\documentclass[iop]{emulateapj}

\newcommand{\kms}{\textrm{km~s$^{-1}$}}

\newcommand{\ergs}{\textrm{erg~s$^{-1}$}}
\newcommand{\lsolar}{L$_{\odot}$}
\newcommand{\msolar}{M$_{\odot}$}
\newcommand{\ml}{M$_{\odot}$ yr$^{-1}$}
\newcommand{\mdot}{\dot{M}}

\usepackage{subfigure}
\usepackage{multirow}
\usepackage[dvips]{color}
\definecolor{Mygrey}{gray}{0.6}

\begin{document}

\title{Late-Time Circumstellar Interaction in a {\it Spitzer}~\\Selected Sample of Type IIn Supernovae}
\shorttitle{SNe Type IIn CSM Interaction}
\author{Ori D. Fox\altaffilmark{1,2}, Alexei V. Filippenko\altaffilmark{1}, Michael F. Skrutskie\altaffilmark{3}, Jeffrey M. Silverman\altaffilmark{1,4}, Mohan Ganeshalingam\altaffilmark{1}, S. Bradley Cenko\altaffilmark{1}, and Kelsey I. Clubb\altaffilmark{1}}
\altaffiltext{1}{Department of Astronomy, University of California, Berkeley, CA 94720-3411.}
\altaffiltext{2}{email: ofox@berkeley.edu}
\altaffiltext{3}{Department of Astronomy, University of Virginia, Charlottesville, VA 22903.}
\altaffiltext{4}{Department of Astronomy, University of Texas, Austin, TX 78712}

\begin{abstract}

Type IIn supernovae (SNe~IIn) are a rare ($< 10$\%) subclass of core-collapse SNe that exhibit relatively narrow emission lines from a dense, pre-existing circumstellar medium (CSM).  In 2009, a warm {\it Spitzer Space Telescope} survey observed 30 SNe~IIn discovered in 2003--2008 and detected 10 SNe at distances out to 175 Mpc with unreported late-time infrared emission, in some cases more than 5 years post-discovery.  For this single epoch of data, the warm-dust parameters suggest the presence of a radiative heating source consisting of optical and X-ray emission continuously generated by ongoing CSM interaction.  Here we present multi-wavelength follow-up observations of this sample of 10 SNe~IIn and the well-studied Type IIn SN 2010jl.  A recent epoch of {\it Spitzer} observations reveals ongoing mid-infrared emission from nine of the SNe in this sample.  We also detect three of the SNe in archival {\it WISE} data, in addition to SNe 1987A, 2004dj, and 2008iy.  For at least five of the SNe in the sample, optical and/or X-ray emission confirms the presence of radiative emission from ongoing CSM interaction.  The two {\it Spitzer} nondetections are consistent with the forward shock overrunning and destroying the dust shell, a result that places upper limits on the dust-shell size.  The optical and infrared observations confirm the radiative heating model and constrain a number of model parameters, including progenitor mass-loss characteristics.  All of the SNe in this sample experienced an outburst on the order of tens to hundreds of years prior to the SN explosion followed by periods of less intense mass loss.  Although all evidence points to massive progenitors, the variation in the data highlights the diversity in SN~IIn progenitor evolution.  While these observations do not identify a particular progenitor system, they demonstrate that future, coordinated, multi-wavelength campaigns can constrain theoretical mass-loss models.

\end{abstract}


\clearpage

\section{Introduction}
\label{sec:intro}

\begin{deluxetable*}{ l c c c c c c c c}
\tablewidth{0pt}
\tabletypesize{\footnotesize}
\tablecaption{$Spitzer$~Observations\tablenotemark{1}\label{tab1}}
\tablecolumns{7}
\tablehead{
\colhead{SN} & \colhead{JD} & \colhead{Epoch} & \colhead{$\alpha$ (h:m:s)} & \colhead{$\delta$ ($^\circ$:$'$:$''$)} & \colhead{Distance\tablenotemark{2}} & \colhead{$t_{\rm int}$} & \colhead{3.6~\micron\tablenotemark{3}} & \colhead{4.5~\micron\tablenotemark{3}}\\
\colhead{}&\colhead{$-$2,450,000}&\colhead{(days)}&\colhead{(J2000)}&\colhead{(J2000)}&\colhead{(Mpc)}&\colhead{(s)} & \multicolumn{2}{c}{(10$^{17}$ erg s$^{-1}$ cm$^{-2}$ \AA$^{-1}$)}
}
\startdata
2002bu & 5734 & 3372 & 12:17:37.18 & +45:38:47.4 & 6.9 & 600 & $<$0.11 & $<$0.09\\
2003lo & 5865 & 2860 & 03:37:05.12 & $-$05:02:17.3 & 57 & 600 & $<$0.11 & $<$0.09\\
2005cp & 5797 & 2254 & 23:59:30.88 & +18:12:09.6 & 91 & 300 & 0.23(0.07) & 0.15(0.05)\\
2005gn & 5875 & 2219 & 05:48:49.07 & $-$24:22:45.5 & 165 & 600 & $<$0.11 & $<$0.09\\
2005ip & 5737 & 2057 & 09:32:06.42 & +08:26:44.4 & 31 & 300 & 7.27(0.27) & 7.64(0.43)\\
2006jd & 5735 & 1714 & 08:02:07.43 & +00:48:31.5 & 77 & 300 & 2.68(0.16) & 2.66(0.24)\\
2006qq & 5875 & 1805 & 05:19:50.43 & $-$20:58:06.4 & 119 & 300 & 0.44(0.10) & 0.28(0.05)\\
2007rt & 5733 & 1304 & 11:02:34.29 & +50:34:58.5 & 93 & 300 & 1.63(0.14) & 1.52(0.12)\\
2008J & 5842 & 1361 & 02:34:24.20 & $-$10:50:38.5 & 66 & 300 & $<$0.11 & $<$0.09\\
2008cg & 5818 & 1226 & 15:54:15.15 & +10:58:25.0 & 151 & 600 & 0.11(0.04) & 0.10(0.03)\\
2008en & 5813 & 1131 & 00:55:13.56 & +35:26:26.2 & 151 & 600 & 0.21(0.05) & 0.19(0.04)\\
2008gm & 5797 & 1035 & 23:14:12.39 & $-$02:46:52.4 & 49 & 300 & 0.14 (0.05) & 0.07 (0.03)\\
2010jl & 5733 & 230 & 09:42:53.33 & +09:29:41.8 & 50 & 600 & 13.08(0.40) & 8.59(0.40)
\enddata
\tablenotetext{1}{Upper limits for nondetections were derived by the point-source sensitivity in Table 2.10 of the IRAC Instrument Handbook, version 2.}
\tablenotetext{2}{Aside from SN 2002bu, all distances are derived from the host-galaxy redshift assuming $H_0$ = 72~km s$^{-1}$~Mpc$^{-1}$.}
\tablenotetext{3}{1$\sigma$~uncertainties are given in parentheses.}
\end{deluxetable*}

Type IIn supernovae (SNe~IIn; \citealt{schlegel90}, see \citealt{filippenko97} for a review) have gained considerable attention over the past decade given their association with explosions ranging from underluminous ``supernova impostors'' \citep[e.g.,][]{smith11impostor,kochanek11,dyk12} to superluminous Type II SNe \citep[SLSNe; e.g.,][]{smith07gy,ofek07}.  Representing {\it fewer} than 10\% of all core-collapse events in the nearby Universe ($d < 60$ Mpc; \citealt{li11}), SNe~IIn are characterized by relatively narrow lines \citep[$\sim 100$--1000 \kms;][]{schlegel90}.  This defining characteristic is not associated with the SN explosion itself, but rather with a dense circumstellar medium (CSM) produced by pre-SN mass loss (though in some cases much of the line width can be attributed to interaction with the SN ejecta).

Identification of a single progenitor class remains ambiguous.  A popular viewpoint connects some high mass-loss properties (e.g., $\mdot \approx 10^{-1}~{\rm to}~10^{-3}$~\ml) to episodic dense winds observed in {\it some} massive stars ($\sim$20--40~\msolar~or greater), including a small fraction of red supergiants \citep[RSGs;][]{fransson02,smith09rsg} and luminous blue variables \citep[LBVs; e.g.,][and references therein]{smith11impostor}.  \citet{gal-yam07} and \citet{gal-yam09} also show direct observational evidence for a LBV progenitor in the case of SN 2005gl.  

The high mass loss, however, is not typical of most LBVs, which achieve rates of only $\mdot < 10^{-4}$~\ml~in their S Doradus state \citep{humphreys94}.  Furthermore, stellar evolution models are inconsistent with LBV progenitors.  Classical stellar evolution theory doesn't even allow for terminal explosions in the LBV state, requiring them to first evolve into a Wolf-Rayet phase \citep{abbott87}.  The models also have difficulty achieving such substantial mass loss.  At best, the mass-loss calculations rely on observed rates or scaled line-driven wind models \citep{smith06}.  Although some recent scenarios present promising leads \citep[e.g.,][]{quataert12}, these models are incomplete and cannot explain why only a fraction of massive stars undergo such extreme mass loss.

While the exact mass-loss mechanism remains unknown, the pre-SN mass-loss history offers an important constraint on different progenitor models since varying wind speeds, densities, and asymmetries result in different observational behaviors.  For example, a dense and optically thick CSM wind will increase the radius of shock breakout relative to the stellar surface \citep[e.g.,][]{moriya11,chevalier11}.  \citet{ofek12} show the relationship between mass loss and the timescale for the shock breakout.  \citet{moriya12} further use nonsteady mass loss to explain the spectral diversity of Type II SLSNe (i.e., those with H in their spectra).  While several core-collapse SNe (CCSNe) not classified as SNe IIn have shown evidence for late-time CSM interaction \citep[e.g.,][]{andrews10,milisavljevic12}, the dense CSM associated with SNe~IIn makes them a particularly good sample for late-time studies ($> 100$ d) to trace the complete mass-loss history of the progenitor.  

Estimated to have a rate of no more than 10 yr$^{-1}$~out to 150 Mpc \citep{dahlen99}, newly discovered SNe~IIn are not easy targets.  Instead, \citet{fox11} executed a warm {\it Spitzer Space Telescope (Spitzer)} survey (P60122) that included 30 SNe~IIn previously discovered in 2003--2008.  The survey, along with observations of SN 2005ip \citep{fox09, fox10}, revealed 10 SNe with previously unreported late-time infrared (IR) excesses, in some cases more than 5 yr post-discovery.  The data are most consistent with pre-existing dust shells with radii $\gtrsim$0.01 pc, formed by a progenitor outburst $\gtrsim$10--100 yr prior to the SN.  The outbursts had mass-loss rates of $\mdot \approx 10^{-1}~{\rm to}~10^{-3}$~\ml~and large total masses $M_{\rm tot} \approx 1$--10 \msolar.  The dust is most likely continuously heated by visible and X-ray radiation generated by ongoing CSM interaction, although direct evidence for such shock interaction in this sample exists only for SNe 2005ip, 2006jd, and 2007rt \citep{smith09ip, fox11,stritzinger12,trundle09}.

X-ray and optical observations of the shock interaction not only validate this model but also probe the inner $< 0.01$ pc of the CSM, which corresponds to mass loss $< 100$ yr prior to the SN.  This paper presents a multi-wavelength follow-up study of the 10 SNe observed by the initial $Spitzer$~survey (as well as the well-studied Type IIn SN 2010jl).  Data include an additional epoch of warm $Spitzer$ photometry, photometry from the {\it Wide-field Infrared Survey Explorer (WISE)}, and ground-based optical photometry and spectroscopy.  The spectral energy distribution (SED) provides a more complete picture of the evolution of SNe~IIn as a forward shock continues to pass through a dense CSM.  Section \ref{sec:2} lists the details of our observations.  In \S \ref{sec:3} we analyze the data with respect to the interaction model.  Implications on the progenitor evolution models are discussed in \S \ref{sec:4}.  Section \ref{sec:5} summarizes the results.

\section{Observations}
\label{sec:2}

\subsection{Warm $Spitzer$/IRAC Photometry}
\label{sec:irac}

Warm $Spitzer$/IRAC \citep{fazio04} obtained a second epoch of data for the 10 SNe~IIn discovered to have late-time IR emission by \citet{fox11} and the well-studied Type IIn SNe 2002bu, 2003lo, and 2010jl (Program PID 80023).  Table \ref{tab1} lists the observational details.  The {\it Spitzer} Heritage Archive\footnote{SHA can be accessed from http://sha.ipac.caltech.edu/applications/Spitzer/SHA/.} provided access to the Post Basic Calibrated Data ({\tt pbcd}), which are already fully coadded and calibrated.  The background flux in most of the SN host galaxies is bright and exhibits rapid spatial variations.  Although template subtraction is a commonly used technique to minimize photometric confusion from the underlying galaxy, no pre-SN observations exist.  Instead, photometry was performed using the {\tt DAOPHOT} point-spread function (PSF) photometry package in {\tt IRAF} (see Figure \ref{f1}).\footnote{IRAF: the Image Reduction and Analysis Facility is distributed by the National Optical Astronomy Observatory, which is operated by the Association of Universities for Research in Astronomy (AURA) under cooperative agreement with the National Science Foundation (NSF).}

\begin{deluxetable}{ l c c c c }
\tablewidth{0pt}
\tabletypesize{\footnotesize}
\tablecaption{$WISE$~Photometry\tablenotemark{1} \label{tab2}}
\tablecolumns{4}
\tablehead{
\colhead{SN} &  \colhead{JD} &  \colhead{Epoch} & \colhead{W1\tablenotemark{2}} & \colhead{W2\tablenotemark{2}}\\
\colhead{ } & \colhead{$-$2,450,000} & \colhead{(days)} & \multicolumn{2}{c}{(10$^{17}$ erg s$^{-1}$ cm$^{-2}$ \AA$^{-1}$)}
}
\startdata
1987A & 5311 & 8461 & 3.12(0.03) & 2.98(0.02) \\
2002bu & 5232 & 2871 & $<$0.27 & $<$0.23\\
2003lo & 5345 & 2341 & $<$0.27 & $<$0.23\\
2004dj & 5285 & 2068 & 0.34(0.06) & 0.51(0.03) \\
2005cp & 5375 & 1833 &  $<$0.27 & $<$0.23\\
2005gn & 5267 & 1612 &  $<$0.27 & $<$0.23\\
2005kd & 5258 & 1572 &  $<$0.27& $<$0.23 \\
2006gy & 5238 & 1242 &  $<$0.27& $<$0.23 \\
2006jd & 5303 & 1283 & 2.79(0.04) & 2.43(0.02) \\
2006qq & 5244 & 1175 & $<$0.27 & $<$0.23\\
2007rt & 5325 & 897 & 5.98(0.25) & 2.66(0.07) \\
2008J & 5400 & 920 &  $<$0.27 & $<$0.23\\
2008en & 5396 & 715 & 0.37(0.03) & 0.23(0.02) \\
2008gm & 5356 & 595 & $<$0.27 & $<$0.23\\
2008iy & 5244 & 507 & 2.42(0.04) & 1.42(0.03) 
\enddata
\tablenotetext{1}{Upper limits for nondetections ($\sim$0.11 mJy at 4.6 \micron) were set by the WISE point-source sensitivity \citep{wright10}.}
\tablenotetext{2}{1$\sigma$ uncertainties are given in parentheses.\\}
\end{deluxetable}

\subsection{WISE Photometry}
\label{sec:wise}

{\it WISE} mapped the entire sky during 2010 at 3.4, 4.6, 12, and 22 \micron~\citep{wright10}.  {\it WISE} therefore included the same SNe that were observed by {\it Spitzer}.  The NASA/IPAC Infrared Science Archive\footnote{IRSA can be accessed from http://irsa.ipac.caltech.edu.} provided fully calibrated images of each field.  Like the {\it Spitzer} data described above, photometry was performed using the {\tt DAOPHOT} PSF photometry package in {\tt IRAF}.  Pixel fluxes were converted from MJy sr$^{-1}$ to mJy according to the Explanatory Supplement to the {\it WISE} All-Sky Data Release Products, which discusses the pixel size, zero points, and aperture correction in detail.  Poor resolution and sensitivity resulted in nondetections for all SNe in the 12 and 22 \micron~bands, so we do not include that photometry here.  Table \ref{tab2} lists the fluxes and upper limits.

\subsection{Optical Photometry and Spectroscopy}
\label{sec:opt}

Tables \ref{tab3} and \ref{tab4} and Figures \ref{f2} and \ref{f3} summarize the optical photometry and spectra.  Data were obtained with the dual-arm Low Resolution Imaging Spectrometer \citep[LRIS;][]{oke95} mounted on the 10-m Keck~I telescope and the DEep Imaging Multi-Object Spectrograph \citep[DEIMOS;][]{faber03}~mounted on the 10-m Keck~II telescope.  Keck/LRIS spectra were obtained using the 600/4000 or 400/3400 grisms on the blue side and the 400/8500 grating on the red side, along with a 1\arcsec~wide slit. This resulted in a wavelength coverage of 3200--9200~\AA\ and a typical resolution of 5--7~\AA.  Keck/DEIMOS spectra were obtained using the 1200/7500 grating, along with a 1\arcsec~wide slit.  This resulted in a wavelength coverage of  4750--7400~\AA\ and a typical resolution of ~3~\AA.  Most observations had the slit aligned along the parallactic angle to minimize differential light losses \citep{filippenko82}; moreover, LRIS is equipped with an atmospheric dispersion corrector.

The photometric images were reduced using standard CCD processing techniques in IDL and Python, utilizing online astrometry programs {\tt SExtractor} and {\tt SWarp}\footnote{SExtractor and SWarp can be accessed from http://www.astromatic.net/software.} (Perley 2012, private communication).  Photometry was performed using the {\tt DAOPHOT} PSF photometry package in {\tt IRAF}.  Calibration was performed using field stars with reported fluxes in the Sloan Digital Sky Survey (SDSS) Data Release 9 Catalogue \citep{ahn12}.  The spectra were reduced using standard techniques \citep[e.g.,][]{foley03,silverman12}. Routine CCD processing and spectrum extraction were completed with {\tt IRAF}, and the data were extracted with the optimal algorithm of \citet{horne86}. We obtained the wavelength scale from low-order polynomial fits to calibration-lamp spectra. Small wavelength shifts were then applied to the data after cross-correlating a template sky to an extracted night-sky spectrum. Using our own IDL routines, we fit a spectrophotometric standard-star spectrum to the data in order to flux calibrate the SN and to remove telluric absorption lines \citep{wade88,matheson00}.  

\begin{figure}[t]
\begin{center}
\epsscale{1.15}
\plotone{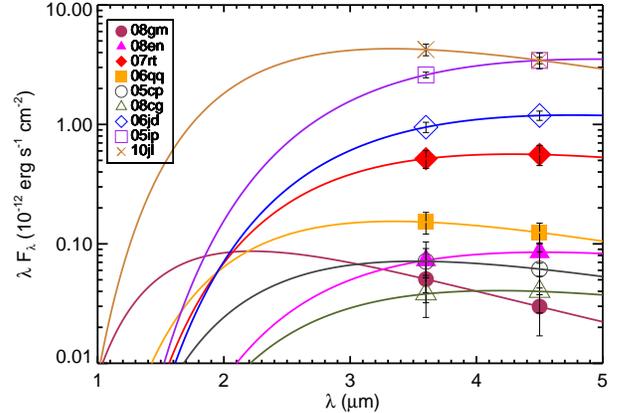}
\caption{Photometry of the SNe detected by $Spitzer$/IRAC bands 1 (3.6~\micron) and 2 (4.5 \micron).  Overplotted are the resulting best fits of Equation \ref{eqn:flux2}.
}
\label{f1}
\end{center}
\end{figure}

\begin{figure*}
\begin{center}
\epsscale{1}
\plotone{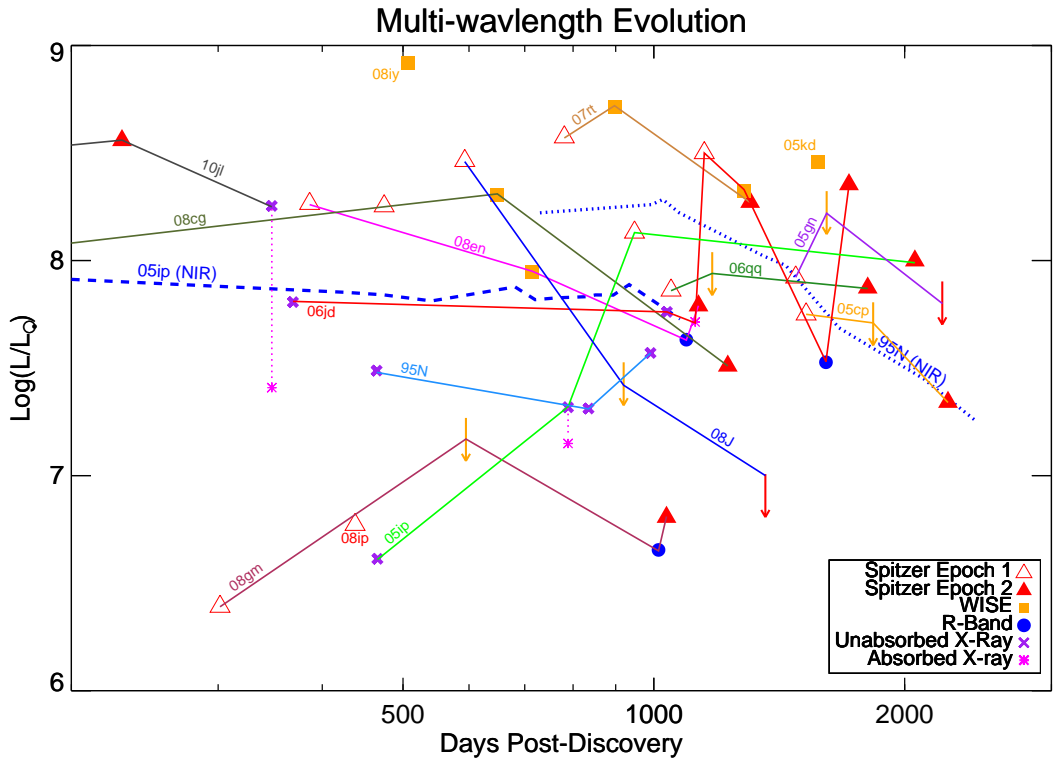}
\caption{The multi-wavelength luminosity evolution of the SNe detected by $Spitzer$.  For reference, the near-IR evolution of two of the most well-studied SNe~IIn (1995N and 2005ip) are overplotted (dashed and dotted lines).  The mid-IR luminosities originate from warm dust, while the optical and X-ray luminosities likely originate from CSM interaction.  While there is a certain degree of scatter in this plot, there is also a noticeable consistency.  The mid-IR luminosities typically fall in the range $10^8 \lesssim L_{\rm mid-IR} \lesssim 10^9$~\lsolar, while the optical and X-ray luminosities are slightly lower.  Described further in the text, these values are consistent with a warm dust shell continuously heated by optical and X-ray emission generated by ongoing CSM interaction.
}
\label{f2}
\end{center}
\end{figure*}

\begin{deluxetable*}{ l c c c c c c c}
\vspace{-0.47in}
\tablewidth{0pt}
\tabletypesize{\footnotesize}
\tablecaption{Keck/LRIS Optical Photometry\tablenotemark{1} \label{tab3}}
\tablecolumns{5}
\tablehead{
\multirow{2}*{SN} &  \colhead{JD} & Epoch & $B$ & $V$ & $R$ & $I$ & Luminosity \\
& \colhead{$-$2,450,000} & (days) & (mag) & (mag) & (mag) & (mag) & (log ($L_{\rm opt}$/\lsolar)) 
}
\startdata 
\multirow{3}*{2005cp} & 5775 & 2233 & $>$22.7 & --- & $>$21.7 & --- & $<$6.7 \\
& 5921 & 2379 & $>$21.7 & $>$22.7 & $>$22.4 & $>$21.6 & $<$6.4 \\
& 6217 & 2675 & $>$22.9 & $>$22.4 & $>$22.4 & $>$21.6 & $<$6.4 \\
\hline
\multirow{2}*{2006jd} & 5629 & 1609 & 22.5 (0.03) & 21.8 (0.03) & 20.4 (0.01) &  21.3 (0.02) & 7.57 \\
& 6046 & 2026 &  $>$23.2 & $>$22.0 & $>$21.8 & $>$21.0 & $<$6.7 \\
\hline
2007rt & 5629 & 1223 & $>$23.3 & $>$22.9 & $>$22.7 & $>$22.3 & $<$6.3 \\
\hline
\multirow{2}*{2008cg} & 5774& 1183  & $>$23.2 & $>$23.1 & $>$22.5 & $>$22.1 & $<$6.4\\
& 6094 & 1503 & $>$22.5 & $>$22.1 & $>$21.7 & $>$21.2 & $<$6.7 \\
\hline
2008en & 5775& 1094 & 22.1 (0.04) & -- & 21.6 (0.03) & 20.9 (0.03) & 7.63 \\
\hline
2008gm & 5774 & 1013 & 21.8 (0.04) & 21.8 (0.03) & 21.6 (0.04) & 20.7 (0.05) & 6.65 
\enddata
\tablenotetext{1}{Upper limits for nondetections reported at the 3$\sigma$~limit.}
\end{deluxetable*}

\begin{deluxetable*}{ l c c c c c c }
\tablewidth{0pt}
\tabletypesize{\footnotesize}
\tablecaption{Summary of Optical Spectra \label{tab4}}
\tablecolumns{6}
\tablehead{
\multirow{2}*{SN} & \colhead{JD} & Epoch & \multirow{2}*{Instrument} & Blue & Red & Int\\
& \colhead{$-$2,450,000} & (days) & & Res. (\AA) & Res. (\AA) & (s)
}
\startdata
\multirow{2}*{2005ip} & 5921 & 2242 & Keck/LRIS & 6.5 & 7 & 900 \\
& 6246 & 2567 & Keck/DEIMOS & -- & 3 & 2400\\
\hline
2005cp & 6246 & 2704 & Keck/DEIMOS & -- & 3 & 2400\\
\hline
\multirow{2}*{2006jd} & 5896 & 1876 & Keck/LRIS & 6.5 & 5.8 & 600 \\
& 6246 & 2226 & Keck/DEIMOS & -- & 3 & 2400\\
\hline
\multirow{2}*{2008en} & 5866 & 1185 & Keck/LRIS & 3.6 & 6.2 & 900\\
& 6193 & 1512 & Keck/DEIMOS & -- & 3 & 2400 \\
\hline
\multirow{2}*{2008gm} & 5866 & 1105 & Keck/LRIS & 3.6 & 6.2 & 900\\
& 6193 & 1432 & Keck/DEIMOS & -- & 3 & 2400 
\enddata
\end{deluxetable*}

\begin{figure*}
\begin{center}
\epsscale{0.7}
\plotone{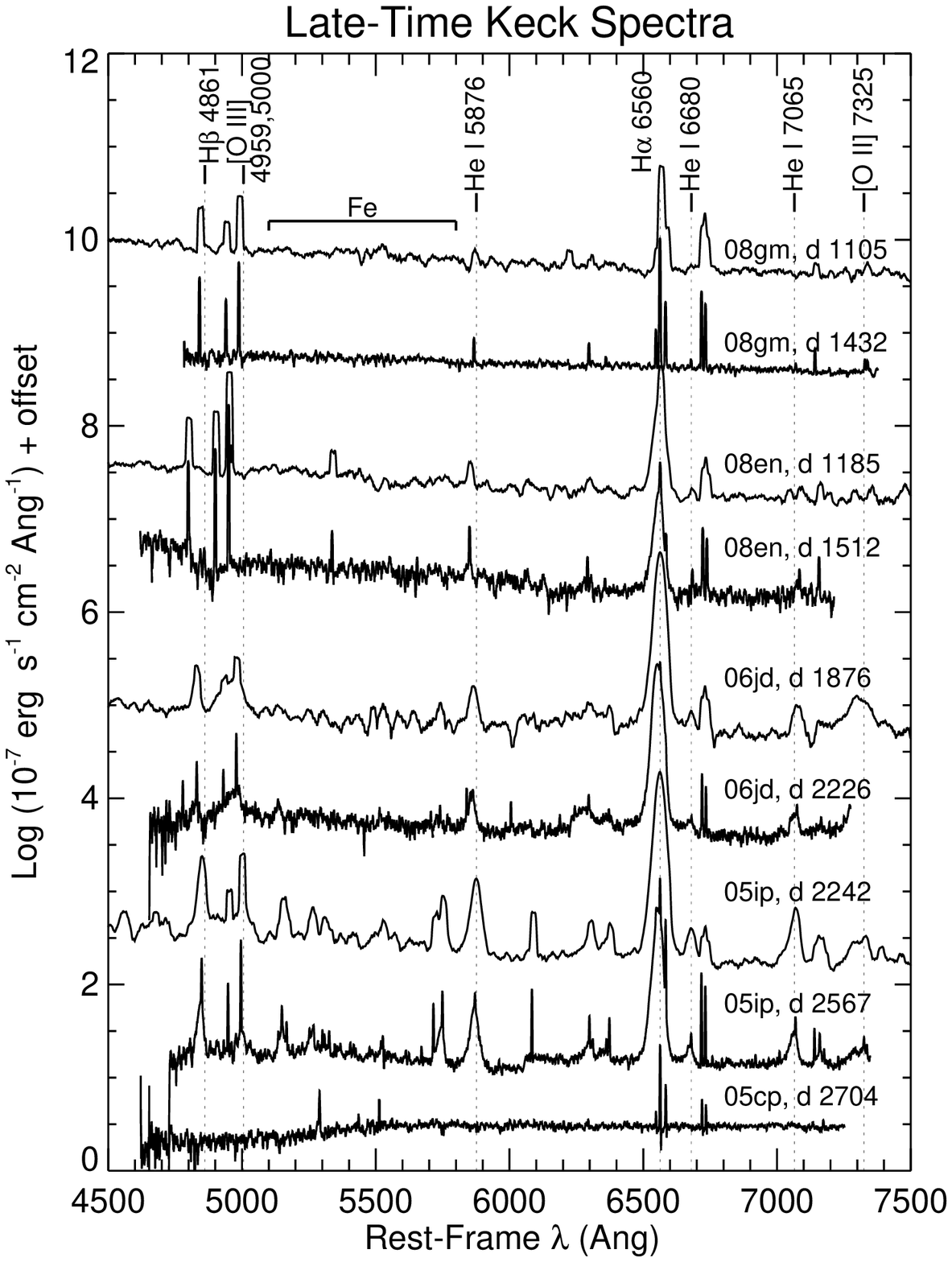}
\caption{Optical spectra of SNe 2005cp, 2005ip, 2006jd, 2008en, and 2008gm taken with Keck/LRIS and DEIMOS.  All spectra were obtained at $> 3$ yr post-explosion.  While the spectra lack any significant continuum, they all show a significant H$\alpha$~line, while SNe 2005ip also exhibits high-ionization Fe lines.
}
\label{f3}
\end{center}
\end{figure*}

\subsection{Dust Temperature and Mass}
\label{sec:dust}

Assuming only thermal emission, the mid-IR photometry provides a strong constraint on the dust temperature, mass, and thus IR luminosity.  Assuming optically thin dust with mass $M_{\rm d}$ and particle radius $a$, at a distance $d$ from the observer, thermally emitting at a single equilibrium temperature $T_{\rm d}$, the flux can be written as \citep[e.g.,][]{hildebrand83}
\begin{equation}
\label{eqn:flux2}
F_\nu = \frac{M_{\rm d} B_\nu(T_{\rm d}) \kappa_\nu(a)}{d^2},
\end{equation}
where $B_\nu(T_{\rm d})$~is the Planck blackbody function and $\kappa_\nu(a)$ is the dust mass absorption coefficient.  

For simple dust populations of a single size composed entirely of either silicate or graphite, the IDL {\tt MPFIT} function \citep{markwardt09} finds the best fit (see Figure \ref{f1}) of Equation \ref{eqn:flux2} by varying $M_{\rm d}$ and $T_{\rm d}$ to minimize the value of $\chi^2$.  The absorption coefficients, $\kappa$, are given in Figure 4 of \citet{fox10}.  With only two mid-IR fluxes, our fits are limited to a single component.  Table \ref{tab5} lists the best-fit parameters for graphite grains of size $a = 0.1$ \micron~(found to be the best-fit value for these SNe by \citealt{fox11}).  Figure \ref{f2} plots the corresponding IR luminosity evolution for each SN, as well as the luminosities derived from {\it R}-band and X-ray observations.

\section{Analysis}
\label{sec:3}

\citet{fox11} show that the dust properties for most of the 10 SNe~IIn in their sample are consistent with a pre-existing dust shell formed in a progenitor wind/eruption and continuously heated by visible, ultraviolet, and/or X-ray emission generated by ongoing CSM interaction in the forward shock.  This scenario yields two testable predictions:  (1) Assuming a single-shell model (${\Delta r}/r = 1/10$), the grains will be sputtered and destroyed when the forward shock reaches the pre-existing dust shell, resulting in a decline in the IR light curve.  If the CSM density drops sufficiently, the degree of shock interaction and, thereby, the optical light curve will also decrease.  A declining IR light curve constrains the dust-shell radius and thus the heating model.  (More complex CSM geometries, such as multiple dust shells \citep[e.g., SN 2009ip;][]{mauerhan12}, may complicate such an analysis.)  (2) Assuming an optically thin dust shell, the observed dust temperature, $T_{\rm d}$, and shell radius, $r_{\rm d}$, require an optical, ultraviolet, and/or X-ray flux given by
\begin{equation}
\label{eqn:lbol}
L_{\rm opt/X}  = \frac{64}{3} \rho a r_{\rm d}^2 \sigma T_{\rm SN}^4 \frac{\int{B_\nu (T_{\rm d}) \kappa(\nu) d\nu}}{\int{B_\nu(T_{\rm SN}) Q_{\rm abs}(\nu) d\nu}} 
\end{equation}
for a dust bulk (volume) density $\rho$ and an effective SN blackbody temperature $T_{\rm SN}$, where $B_\nu$~is the Planck blackbody function, $Q_{\rm abs}$ is the dust absorption efficiency, and $\kappa(\nu)$ is the dust mass absorption coefficient.  Here we test these predictions.

\begin{deluxetable}{ l l c c c }
\tablewidth{0pt}
\tabletypesize{\footnotesize}
\tablecaption{IR~Fitting Parameters ($a = 0.1$ \micron) \label{tab5}}
\label{tab_phot}
\tablecolumns{5}
\tablehead{
\colhead{SN} & \colhead{Telescope} & \colhead{$T_{\rm d}$} (K) & \colhead{$M_{\rm d}$ (\msolar)} & \colhead{$L_{\rm d}$ (\lsolar)}
}
\startdata
2004dj & $WISE$ &  486 & 2.39$\times10^{-5}$ & 9.19$\times10^{4}$ \\
2005cp & $Spitzer$ & 734 & 5.17$\times10^{-4}$ & 2.20$\times10^{7}$ \\
2005ip & $Spitzer$ & 511 & 1.93$\times10^{-2}$ & 9.98$\times10^{7}$ \\
2005kd & $WISE$ & 795 & 4.23$\times10^{-3}$ & 2.87$\times10^{8}$ \\
2006gy & $WISE$ & 1287 & 5.57$\times10^{-3}$ & 6.52$\times10^{9}$ \\
2006jd & $Spitzer$ & 531 & 3.49$\times10^{-2}$ & 2.26$\times10^{8}$ \\
2006jd & $WISE$ & 574 & 2.08$\times10^{-2}$ & 2.11$\times10^{8}$ \\
2006qq & $Spitzer$ & 771 & 1.32$\times10^{-3}$ & 7.47$\times10^{7}$ \\
2007rt & $Spitzer$ & 555 & 2.25$\times10^{-2}$ & 1.87$\times10^{8}$ \\
2007rt & $WISE$ & 834 & 5.80$\times10^{-3}$ & 5.19$\times10^{8}$ \\
2008cg & $Spitzer$ & 575 & 3.14$\times10^{-3}$ & 3.24$\times10^{7}$ \\
2008cg & $WISE$ & 807 & 2.74$\times10^{-3}$ & 2.02$\times10^{8}$ \\
2008en & $Spitzer$ & 563 & 6.77$\times10^{-3}$ & 6.17$\times10^{7}$ \\
2008en & $WISE$ & 706 & 2.63$\times10^{-3}$ & 8.89$\times10^{7}$ \\
2008gm & $Spitzer$ & 1074 & 1.61$\times10^{-5}$ & 6.41$\times10^{6}$ \\
2008iy & $WISE$ & 730 & 2.01$\times10^{-2}$ & 8.29$\times10^{8}$ \\
2010jl & $Spitzer$ & 735 & 8.47$\times10^{-3}$ & 3.62$\times10^{8}$
\enddata
\end{deluxetable}

\subsection{Dust-Shell Destruction}
\label{sec:destruction}

The forward shock will eventually destroy the dust via sputtering, resulting in a declining mid-IR component.  For a spherically symmetric dust shell, the shock will reach the dust shell at a time $t \gtrsim r_{\rm bb}/v_{\rm s}$, where $v_{\rm s}$~is the forward-shock velocity and $r_{\rm bb}$~is the blackbody radius, which is derived from the IR luminosity in Table \ref{tab5} assuming an optically thick shell.  This equation places a lower limit on the timescale over which dust destruction can occur.  For example, the forward shock may decelerate with time.  Also, an optically thin dust shell composed of smaller grains emitting with modified blackbodies requires a larger shell radius to generate the observed flux.  Both of these effects increase the time over which it takes the forward shock to reach the dust shell.

\begin{figure}
\epsscale{1.15}
\begin{center}
\plotone{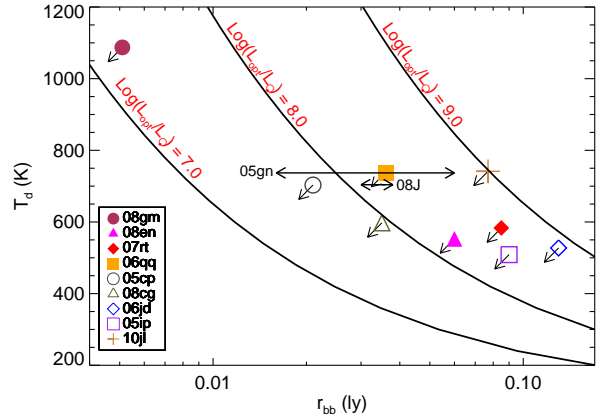}
\caption{Plot of the {\it observed} dust temperature, $T_{\rm d}$,~and blackbody radius, r$_{\rm bb}$.  The diagonal arrows signify that these values reflect upper limits given the limitations of the single-component fit in Figure \ref{f1}.  The horizontal arrows for SNe 2005gn and 2008gn signify both the lower and upper limits on the dust-shell radius derived from the data.  Overplotted are functions of the {\it theoretical}~dust temperature as a function of radius for contours of constant optical/X-ray luminosities.  The results highlight that, in almost every case, the warm dust can be powered by optical and/or X-ray luminosities of  $10^8 \lesssim L_{\rm opt/X} \lesssim 10^9$~\lsolar.  Accounting for absorption, these luminosities are consistent with the observations shown in Figure \ref{f2} and those reported by \citet{smith09ip} and \citet{chandra12jd}.
}
\label{f4}
\end{center}
\end{figure}

For each SN detected by $Spitzer$, Figure \ref{f4} plots the blackbody radius and dust temperature.  The diagonal arrows signify that these values are upper limits given the limitations of the single-component fit in Figure \ref{f1}.  While the numbers vary slightly, a blackbody radius $r_{\rm bb} \lesssim 0.1$ ly and a shock velocity $v_{\rm s} \approx 3000$ \kms~will result in a mid-IR plateau that continues for roughly $t \lesssim 10.5$ yr.  Once the shock reaches the shell, the luminosity decline will occur on timescales consistent with the dust-shell radius ($t_{\rm decline} = 2 r_{\rm bb}/c \approx 0.2$~yr), as the dust destruction on the near side of the shell will be observed before dust destruction on the far side.

We search for evidence of dust destruction between the first and second epochs of $Spitzer$~observations.  Figure \ref{f2} shows that SNe 2005gn and 2008J are not detected on the second epoch of observations, which occurred at 6.1 and 3.7 yr post-explosion, respectively.  Assuming a 3000 \kms~shock, these results place upper limits on the dust-shell radii of $r_{\rm d} < $0.06 and 0.036 ly.  A slower shock would result in a proportionally smaller shell radius.  While still detected, SN 2005cp also shows evidence of fading nearly 6.2 yr post-explosion.  The horizontal arrows in Figure \ref{f4} illustrate both the lower and upper limits on the dust-shell radius derived for SNe 2005gn and 2008J.
\vspace{0.2in}

\subsection{Shock Interaction and CSM Characteristics}
\label{sec:interaction}

For the scenario in which the dust shell is heated by continuous radiation generated by ongoing CSM interaction, an optical and/or X-ray counterpart should exist.  Equation \ref{eqn:lbol} calculates this luminosity as a function of the observed dust temperature, $T_{\rm d}$, and shell radius, $r_{\rm d}$.  Along with the {\it observed} dust temperatures and radii, Figure \ref{f4} plots the {\it theoretical} dust temperature as a function of shell radius for contours of constant luminosity.  In most cases, the observed dust parameters are consistent with optical and/or X-ray luminosities $10^8 \lesssim L_{\rm opt/X} \lesssim 10^9$~\lsolar.

Figure \ref{f2} plots the observed optical and X-ray luminosities for each SN.  For SNe 2008gm and 2008en, the optical and/or X-ray luminosities are fairly consistent with the contours in Figure \ref{f4}.  The optical and/or X-ray luminosities may seem a little low compared to the expected values given by the dust parameters in Figure \ref{f4}, but we stress that the observed dust parameters are only upper limits due to the limitations of a single-component fit.  Furthermore, even in an optically thin scenario (i.e., $\tau < 1$), 50\% of the optical flux can be absorbed by the dust.  Overall, the optical/X-ray fluxes are sufficient to power the mid-IR emission.  Although Figure \ref{f2} excludes previously published optical data, observations by \citet{smith09ip}, \citet{stritzinger12}, and \citet{zhang12} yield similar conclusions for SNe 2005ip, 2006jd, and 2010jl, respectively.  Due to the high background flux from their underlying galaxies, upper limits for SNe 2005cp, 2007rt, and 2008cg (see Table \ref{tab3}) do not rule out the presence of high-energy emission from CSM interaction.  No observations were acquired of SN 2006qq.

\begin{figure}
\epsscale{1.15}
\begin{center}
\plotone{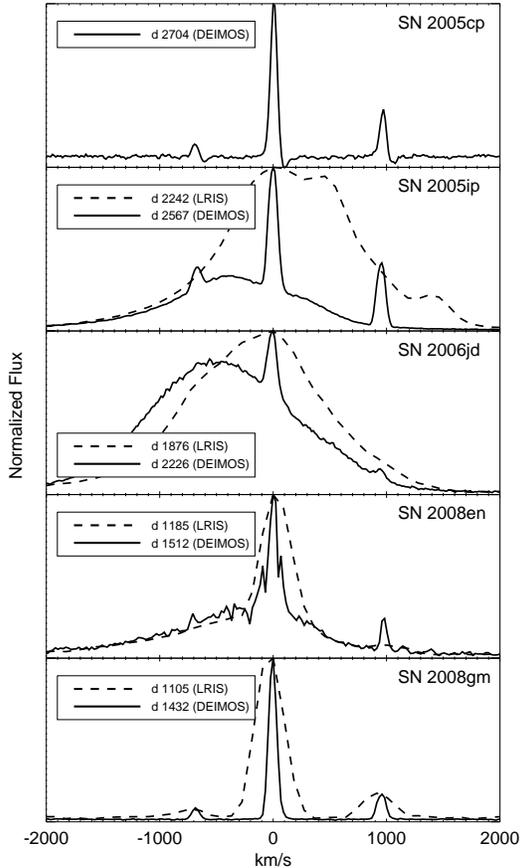}
\caption{Plots of the H$\alpha$~line for the spectra in Figure \ref{f3}.  These plots show both low- and moderate-resolution spectra obtained with Keck LRIS and DEIMOS, respectively.  A post-shocked CSM will typically have velocities $> 100$ \kms.  These relatively broad lines are prominent in the LRIS spectra of SNe 2005ip and 2006jd, but more ambiguous in the other LRIS spectra.  The moderate-resolution DEIMOS spectra differentiate between the broader component of CSM interaction (in the case of SN 2008en) and the narrower ($< 100$ \kms) component likely associated with an underlying H~II region (SNe 2005cp and 2008gm).  The blueshift (relative to the rest wavelength) of the broader component is likely due to dust absorption \citep{smith09ip,stritzinger12}. 
}
\label{f5}
\end{center}
\end{figure}

SN 2010jl does stand out, however, with a relatively large dust temperature for the derived blackbody radius.  \citet{andrews11} derive similar parameters from $Spitzer$~observations obtained on day 90 post-explosion and suggest that these characteristics are more consistent with an IR echo powered by the initial SN flash.  Optical and X-ray observations, however, reveal clear signatures of CSM interaction.  Continued monitoring will ultimately constrain the dust-shell size and heating source.

Although $Spitzer$~did not detect SNe 2005gn and 2008J, the dust shell upper limits (0.6 and 0.036 ly, respectively) derived in \S \ref{sec:destruction} offer valuable constraints on their possible heating mechanism.  Assuming dust temperatures (737 K and 704 K) derived from the first epoch of $Spitzer$~observations by \citet{fox11}, Figure \ref{f4} shows that the dust never required an optical and/or X-ray luminosity greater than $10^{8.5}$~\lsolar.  While we do not have late-time optical observations of these two SNe, such limits are consistent with the other optical and X-ray observations.

The observed optical and X-ray luminosities are consistent with the physical scenario described above, but the source of the emission is not immediately clear.  For example, the emission may originate from ongoing CSM interaction, but it may also be associated with an underlying H~II region.  To identify distinct signatures of CSM interaction, we obtained ground-based optical spectra of the four SNe detected at visual wavelengths (SNe 2005ip, 2006jd, 2008en, and 2008gm) and SN 2005cp, as summarized in Table \ref{tab4} and Figure \ref{f3}. While SN 2005ip exhibits clear signs of CSM interaction via the high-ionization Fe lines and broadened H$\alpha$, the evidence in the other SNe is more ambiguous.  

Unlike the LRIS spectra, the moderate-resolution DEIMOS spectra resolve down to $< 100$ \kms~and can differentiate between H~II regions and CSM interaction, which typically has post-shock velocities $>$100 \kms.  Figure \ref{f5} illustrates these differences with the H$\alpha$~line.  SNe 2005ip and 2006jd both have distinct, relatively broad ($> 1000$ \kms) features associated with the forward shock and a narrower ($\sim 100$--200 \kms) component likely associated with the pre-shocked CSM \citep{fox11}.  The line-profile asymmetry (a more prominent blue side) of the broader component is likely due to dust absorption \citep{smith09ip,stritzinger12}.  The broad feature in SN 2008en is less obvious, particularly in the LRIS spectrum.  The moderate-resolution DEIMOS spectrum, however, resolves the two components.  The LRIS spectrum of SN 2008gm also does not rule out a component of $\sim 500$ \kms.  The DEIMOS spectra of SNe 2005cp and 2008gm, however, show only a narrow ($< 100$ \kms) line suggestive of an H~II region.  

\section{Discussion}
\label{sec:4}

\subsection{Mass-Loss History}
\label{sec:massloss}

The mid-IR traces the characteristics of the CSM at the dust-shell radius.  Assuming a dust-to-gas mass ratio expected in the H-rich envelope of a massive star, $Z_{\rm d} = M_{\rm d}/M_{\rm g} \approx 0.01$, the dust-shell mass can be tied to the progenitor mass-loss rate,
\begin{eqnarray}
\label{eqn:ml}
\mdot & = & \frac{M_{\rm d}}{Z_{\rm d} \Delta r} v_{\rm w} \\ 
& = & \frac{3}{4} \Big(\frac{M_{\rm d}}{\rm M_{\odot}}\Big) \Big(\frac{v_{\rm w}}{120~\rm km~s^{-1}}\Big) \Big(\frac{0.05~\rm ly}{r}\Big) \Big(\frac{r}{\Delta r}\Big) {\rm M}_{\odot}~{\rm yr}^{-1},
\end{eqnarray}
for a progenitor wind speed $v_{\rm w}$.  The relatively narrow lines of SNe~IIn originate in the slow pre-shocked CSM and therefore correspond to the progenitor wind speed. For these SNe, the narrow-line profiles have a full width at half-maximum intensity (FWHM) in the range $\sim 100$--500 \kms \citep{fox11}.  Assuming a thin shell, ${\Delta r}/r = 1/10$, Table \ref{tab6} lists the associated mass-loss rate for each SN.

\begin{figure*}
\begin{center}
\epsscale{1.15}
\plotone{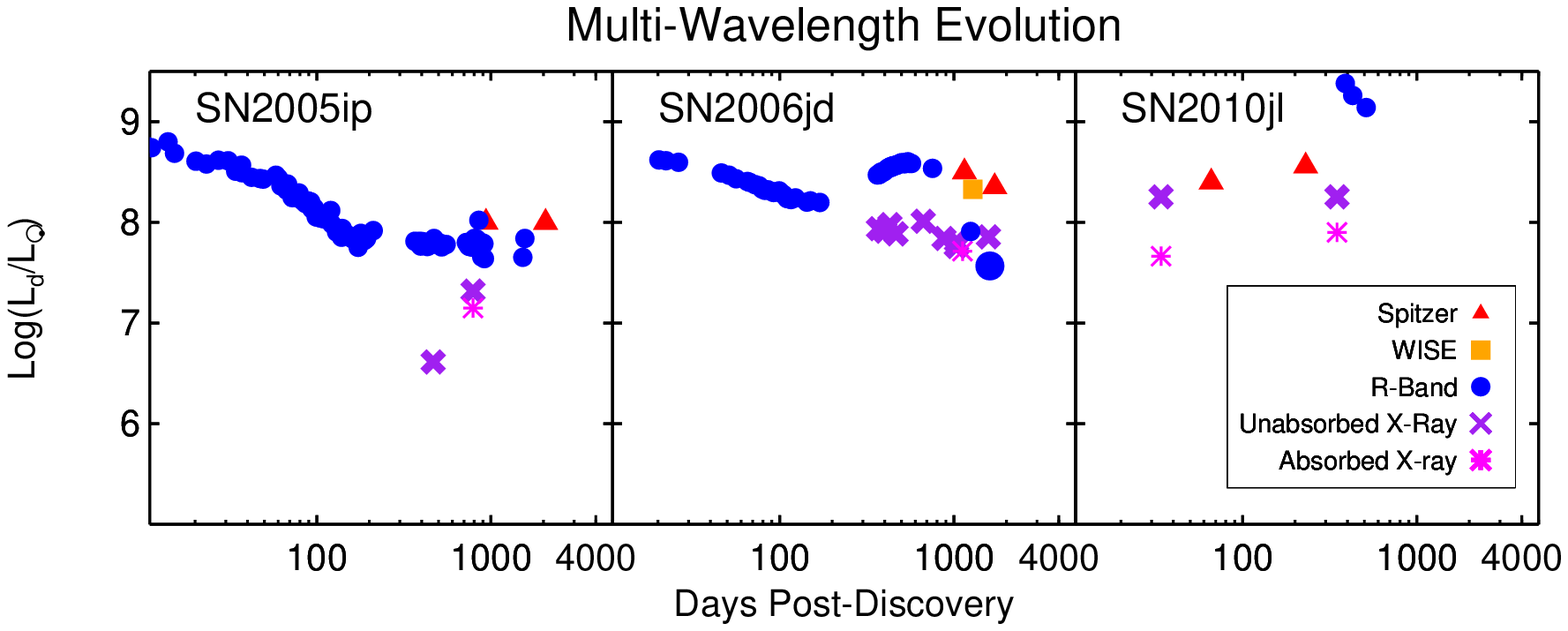}
\caption{Individual plots of the SNe in Figure \ref{f2} having the best coverage.  Also plotted are pre-existing mid-IR, optical, and X-ray photometry \citep{immler07,smith09ip,fox11,andrews11,chandra12jd,chandra12jl,zhang12,stritzinger12}.  While the mid-IR luminosities are comparable, the observations are sparse and comparisons are difficult.  The optical and X-ray evolution, however, show significant contrasts that likely originate in the progenitor.  Multi-wavelength light curves such as these offer useful constraints on theoretical models discussed in the text.
}
\label{f6}
\end{center}
\end{figure*}

 \begin{deluxetable}{ l c c }
\tablewidth{0pt}
\tabletypesize{\footnotesize}
\tablecaption{Mass-Loss Rates}
\label{tab_phot}
\tablecolumns{3}
\tablehead{
\colhead{SN} & \colhead{$\dot{M}$(IR)} & \colhead{$\dot{M}$(opt/X) \label{tab6}}\\
  & (\ml) & (\ml)
}
\startdata
2005cp & 3.4$\times10^{-2}$ & -- \\
2005ip & 8.3$\times10^{-2}$ & 1.7$\times10^{-2}$\\
2006jd & 1.7$\times10^{-1}$ & 2.2$\times10^{-3}$\\
2006qq & 1.2$\times10^{-1}$ & -- \\
2007rt & 1.0$\times10^{-1}$ & -- \\
2008cg & 6.1$\times10^{-2}$ & -- \\
2008en & 7.8$\times10^{-2}$ & 1.1$\times10^{-3}$\\
2008gm & 2.8$\times10^{-3}$ & -- \\
2010jl & 8.8$\times10^{-2}$ & 1.2$\times10^{-4}$
\enddata
\end{deluxetable}

By contrast, the optical and/or X-ray observations trace the CSM characteristics at the inner radii.  Assuming a steady mass loss, the rate can be written as a function of the optical/X-ray luminosity, progenitor wind speed, and shock velocity \citep[e.g.,][]{chugai94, smith09ip}:
\newpage
\begin{eqnarray}
\label{eqn:inner}
\mdot & = & \frac{2 v_w}{\epsilon v_{s}^3}L_{\rm opt/X},\\
 & = & 2.1 \times 10^{-4} \Big(\frac{L_{\rm opt/X}}{3~\times 10^{41}~ \ergs}\Big) \Big(\frac{\epsilon}{0.5}\Big)^{-1} \times \nonumber\\
  & & \Big(\frac{v_w}{120~\kms}\Big) \Big( \frac{v_s}{10^4~\kms}\Big)^{-3} {\rm M_{\odot}~yr^{-1}},
\end{eqnarray}
where $\epsilon < 1$~is the efficiency of converting shock kinetic energy into visual light.  While the conversion efficiency varies greatly depending on shock speed and wind density, we assume $\epsilon \approx 0.5$, acknowledging that this value may be high.  Table \ref{tab6} also lists the mass-loss rates for the SNe at the inner radii based on the optical/X-ray luminosities in Table \ref{tab3}.

For the SNe with both IR and optical observations, Table \ref{tab6} highlights that the dust shells correspond to a period of higher mass loss and densities.  By contrast, the CSM that generates the optical and/or X-ray emission corresponds to a period of lower mass-loss rates.  The progenitor mass loss was variable, not continuous.  The rates given in Table \ref{tab6} (e.g., $\mdot \approx 10^{-1}~{\rm to}~10^{-3}$~\ml) are consistent with previous comparisons of SNe~IIn to the episodic dense winds observed in some massive stars \citep[e.g.,][]{fransson02,smith09rsg,smith11impostor}.  In almost every case, the dust-shell radius and progenitor wind speeds suggest outbursts that occurred on the order of tens to hundreds of years prior to the SN explosion.  

\subsection{Theoretical Models and Progenitor Diversity}
\label{sec:diversity}

While stellar evolution models are still limited in their ability to achieve such substantial mass loss, the nature of the CSM wind can constrain the progenitor mass and explosion physics \citep[e.g.,][]{balberg11}.  The observations in this paper present only a single snapshot of the CSM environment, but a number of theoretical models can generate the entire multi-wavelength evolution as a function of the progenitor mass-loss history.  For example, the early ($< 100$ d) optical and X-ray evolution can be described as a function of the immediate CSM density and explosion energy \citep[e.g.,][]{chevalier11,moriya11, chevalier12, moriya12}.  Furthermore, \citet{ofek12} and \citet{svirski12} show that for a sufficiently dense CSM, the X-rays will escape only months to years {\it after} the SN peaks in the visible.  The escape of the X-rays (sometimes referred to as a circumstellar breakout) will differ observationally for varying progenitor mass-loss rates \citep{chevalier11}.  To fully understand the mass-loss history of the progenitor therefore requires well-sampled multi-wavelength observations.

Figure \ref{f6} plots the multi-wavelength data available for individual SNe in our sample, including data from Figure \ref{f2} and pre-existing optical and X-ray photometry for SNe 2005ip, 2006jd, 2007rt, and 2010jl \citep{immler07,smith09ip,fox11,chandra12jd,chandra12jl,zhang12,stritzinger12}.  While fitting the theoretical models (such as those by \citealt{moriya12} and \citealt{svirski12}) is beyond the scope of this paper, Figure \ref{f6} serves as a potentially useful template for theorists \citep[e.g.,][]{pan13}.  

Furthermore, the figure highlights the diversity among SNe~IIn at various wavelengths.  For example, the mid-IR luminosities are similar and serve as useful indicators for shock interaction and optical/X-ray emission.  The optical and X-ray evolution differ significantly, suggestive of different progenitor mass-loss characteristics and, perhaps, different progenitors.  In the case of SN 2010jl, the absorbed luminosity increases by a factor of 2 more than 1 yr after the visible maximum, revealing the first case of an X-ray breakout from a dense CSM \citep{chandra12jl}.  By contrast, SN 2006jd shows an X-ray plateau but no sign of a shock breakout, while SN 2005ip has insufficient X-ray observations.  The optical light curve of SN 2010jl is quite luminous ($>10^9$~\lsolar), but fades throughout the extent of the observations, whereas SN 2006jd peaks at $\sim$500 d and SN 2005ip plateaus through more than 1000 d.  In the case of SN 2006jd, a large X-ray luminosity implies a high density, but a small amount of photoabsorption in the spectrum implies a low density \citep{chandra12jd}.  These conflicting measurements suggest an asymmetric CSM.  The slow evolution of the unabsorbed emission further suggests a progenitor wind-density profile that does not follow the standard $r^{-2}$~model.  In the case of SN 2005ip, the late-time optical luminosity is over a magnitude larger than the X-ray luminosity.  SN 2005ip does, however, exhibit very high-ionization coronal lines in the visible spectrum.  \citet{smith09ip} explain these characteristics with a lower density CSM interspersed with denser clumps.  The shock interaction with the dense clumps generates X-rays, but with less intensity than in SN 2006jd.  Since the clumps have a small filling factor, the X-rays can escape and photoionize the dense, pre-existing CSM at larger radii.

\section{Conclusion}
\label{sec:5}

This paper presents multi-wavelength follow-up observations of a sample SN~IIn population identified by late-time mid-IR emission from warm dust \citep{fox11}.  For at least five of the nine detected SNe, the data identify the predominant heating source of the dust as radiative optical and/or X-ray emission continuously generated by ongoing CSM interaction.  Optical spectra of SNe 2005ip, 2006jd, 2008en, 2008gm, and 2010jl \citep{smith12jl} confirm the CSM interaction with the presence of relatively broad H$\alpha$~lines.  The optical spectra of SNe 2005cp and 2008gm reveal only narrow lines, suggestive of an H II region, but these spectra were obtained $> 6$ months following the initial $Spitzer$~detection.  The IR emission may have faded, or radiative shock emission does not provide the dominant heating mechanism.  Also, $Spitzer$~did not detect SNe 2005gn or 2008J, consistent with the forward shock overrunning and destroying the dust shell.  The nondetections place upper limits on the dust-shell size, which are also consistent with the radiative shock model despite the lack of shorter wavelength observations.  Further $Spitzer$~observations will constrain the remaining dust-shell radii and resolve some of the ambiguity surrounding SNe 2005cp and 2008gm.

The multi-wavelength data trace the mass-loss history of the progenitor.  All of the SNe in this sample appear to have undergone an outburst on the order of tens to hundreds of years prior to the SN explosion.  More recently, the mass loss appears to be more sporadic.  The multi-wavelength light curves vary significantly, particularly at optical wavelengths, suggestive of a range of Type IIn progenitors.  However, $Spitzer$~did {\it not}~detect 20 of the SNe~IIn that occurred between 2003 and 2008.  From Equation \ref{eqn:inner}, a shock velocity $v_s \approx 10^3$~\kms~into a CSM produced by a steady wind with a mass-loss rate of only $10^{-4}$~\ml~emits a luminosity of only $\sim 10^6$~\lsolar, which would be below the detection threshold for most of these SNe.  Furthermore, lower mass-loss rates also correspond to lower densities, which result in lower dust masses.  These nondetections likely correspond to progenitors that had mass-loss rates $\lesssim 10^{-4}$~\ml, more in line with LBVs in their S Doradus state \citep{humphreys94}.

The diversity amongst this sample parallels the larger SN~IIn population.  The relatively narrow lines and dense CSM associated with SNe~IIn have now been identified in an unexpectedly diverse list of subclasses, including SLSNe \citep[e.g., SN 2006gy;][]{smith07gy}, Type IIP \citep[e.g., SN 2011ht;][]{mauerhan12}, Type IIL \citep[e.g., SN 1998S;][]{liu00,pozzo04}, Type Ibc \citep[e.g., SN 2006jc;][]{foley07,smith08jc}, ``impostors'' \citep{kochanek12}, and even Type Ia \citep[e.g., SN 2002ic, SN 2008J;][]{hamuy03, taddia12, silverman13}.  Existing multi-wavelength data do not come close to sampling the wide diversity of SNe~IIn.  Future, well-sampled, multi-wavelength observations over year-long timescales will be required to classify these different types based on the nature of the X-ray escape time and the precise relationship between CSM interaction and warm dust emission in the mid-IR.

\vspace{10 mm}

We thank the referee, Geoff Clayton, for his helpful comments.  This work is based on observations made with the {\it Spitzer Space Telescope} (PID 80023), which is operated by the Jet Propulsion Laboratory, California Institute of Technology, under a contract with NASA. Support for this work was provided by NASA through an award issued by JPL/Caltech.  Keck/LRIS imaging data were reduced using excellent software provide Daniel Perley. O.D.F. would like to thank conference organizers for useful conversations generated at the Massive Stars Workshop, held at the University of Minnesota in October, 2012.  A.V.F. and his group acknowledge generous financial assistance from Gary and Cynthia Bengier, the Richard and Rhoda Goldman Fund, the Christopher R. Redlich Fund, the TABASGO Foundation, and NSF grant AST-1211916. Some of the data presented herein were obtained at the W. M. Keck Observatory, which is operated as a scientific partnership among the California Institute of Technology, the University of California, and NASA; the observatory was made possible by the generous financial support of the W. M. Keck Foundation. We thank the staff of the Keck Observatory for their assistance with the observations.

\bibliographystyle{apj}
\bibliography{references}

\end{document}